\documentclass[11pt]{article}
\usepackage[margin=0.88in]{geometry}
\usepackage{graphicx}
\usepackage{booktabs}
\usepackage{amsmath}
\usepackage{amssymb}
\usepackage{microtype}
\usepackage{hyperref}
\usepackage{enumitem}
\usepackage{caption}
\usepackage{xcolor}
\usepackage{natbib}
\usepackage{authblk}
\usepackage{array}
\usepackage{multirow}
\usepackage{longtable}
\usepackage{url}
\hypersetup{colorlinks=true,linkcolor=blue,citecolor=blue,urlcolor=blue}
\title{Scoring With the Engine: Retrieval Exposure, Cross-Engine Divergence, and the Limits of Engine-Agnostic GEO Scores}
\author[1]{Benjamin Tannenbaum}
\affil[1]{Aiso Boost Ltd.}
\date{September 2026}

\begin{document}
\maketitle

\begin{abstract}
Recent work asks whether generative-engine visibility can be approximated with deterministic, engine-free page scores. \citet{bajemon2026} provides an unusually careful answer: query-agnostic page scores have weak within-query association with citation order, while query-conditioned relevance is substantially more informative in controlled candidate-set experiments. That design isolates an important quantity, citation preference conditional on exposure, but intentionally removes the production process that determines whether a page is searched, retrieved, reranked, placed in context, and ultimately made available for citation.

We measure that missing production layer using an observational audit of four live AI search surfaces. On 6 June 2026, a fixed benchmark of 15 commercial prompts produced 589 citation observations across ChatGPT, Microsoft Copilot, Google, and Perplexity, corresponding to 528 unique exact URLs and 356 domains. Same-prompt cross-engine URL overlap was extremely small: mean pairwise Jaccard similarity was 0.0079 (prompt-clustered bootstrap 95\% CI 0.0037--0.0132), the median was zero, and 84.9\% of engine pairs shared no cited URL. On the ten prompts observed on all four engines, mean exact-URL Jaccard was 0.0072, reciprocal-rank-weighted overlap was 0.0027, and top-five exact-URL overlap was zero in all 60 pairwise comparisons. A single engine captured only 11.4\%--42.6\% of the four-engine URL union, while the union contained 2.43 times as many URLs as the broadest single engine on average. Among all URLs observed on 6 June, 96.4\% appeared in only one engine. A separate 5-to-6 June same-engine comparison found mean URL-set turnover of 67.0\% (95\% CI 61.1\%--72.2\%). Despite that short-horizon movement, a same-engine adjacent-day citation set was still about 42 times more similar than a same-day cross-engine set in this benchmark.

These results do not invalidate engine-free page scoring. They identify its estimand. A score computed without a live engine can estimate page quality or query-page fit, but end-to-end visibility additionally depends on engine-specific exposure and selection. We formalize this distinction, connect it to recent work on GEO, query rewriting, RAG set selection, source attribution, citation absorption, and generative-search auditing, and propose reporting page fit, observed exposure, conditional selection, and final visibility as separate quantities. The practical implication is that an ``AI visibility score'' that never observes an engine should not be interpreted as a stable probability of being surfaced or cited by that engine without an explicit exposure model.
\end{abstract}

\section{Introduction}
Generative search changes the object being measured. Traditional web search exposes a ranked list of results. A generative search product may decide whether web search is needed, rewrite or fan out the user's request into one or more searches, retrieve a candidate pool, rerank or truncate it, allocate some material to context, generate an answer, and attach citations that need not enumerate every page that influenced the response. Work on retrieval-augmented generation has long separated retrieval from generation \citep{lewis2020,asai2024}; generative-search studies increasingly distinguish discoverability, citation, factual support, and answer influence \citep{liu2023,zhang2026,martinez2026}.

That distinction matters for a fast-growing class of engine-agnostic GEO scores. A deterministic function can inspect a page and estimate whether it is structured, authoritative, fresh, evidence-rich, or semantically aligned with a request. Such a score can be cheap, stable, auditable, and useful. The harder question is what it estimates.

\citet{bajemon2026} directly tests this issue. Their study stress-tests a deterministic content score, re-evaluates several earlier GEO interventions under newer model conditions, identifies and corrects a ranking-evaluation leakage problem, and shows that query-agnostic scoring has only weak within-query predictive power for citation rank. Query-conditioned relevance performs materially better. Importantly, their strongest ranking experiments give the answering model a fixed set of candidate sources. This is the correct design for isolating what happens \emph{after} exposure, but it cannot test the upstream production process that determines which sources are exposed in the first place.

This paper studies that missing layer. We ask a deliberately narrow question: \textbf{when the user request is held fixed, do production generative-search engines expose observably similar citation sets?} If the answer were yes, then a strong query-page fit score could plausibly serve as a broad proxy for end-to-end visibility. If the answer is no, the engine itself remains a material part of the estimand.

Our earlier work motivates this focus on measurement boundaries. \citet{tannenbaum2026arsd} measures how much search work is compressed into conversational answers. \citet{tannenbaum2026prompt} shows that an endpoint prompt often omits request-state dimensions established earlier in a conversation. \citet{tannenbaum2026context} demonstrates experimentally that restoring within-conversation context changes answers materially in a large fraction of cases. \citet{tannenbaum2026purchase} shows a different observability boundary: recommendations are visible in conversation logs far more often than subsequent purchase decisions. The common methodological principle is that an observed artifact is not automatically a measurement of the full process that produced it.

We make six contributions. First, we formalize live citation visibility as a multi-stage exposure-selection process and locate engine-free scoring inside that process. Second, we quantify exact URL and domain divergence across four production AI search surfaces on matched prompts and dates. Third, we add rank-sensitive, top-$k$, source-exclusivity, and proxy-coverage analyses that make the operational consequence of low overlap concrete. Fourth, we measure source breadth and concentration by engine. Fifth, we separate cross-engine divergence from short-horizon temporal turnover and show that the engine effect is much larger than the observed adjacent-day similarity change. Sixth, we synthesize these findings with recent GEO, generative-search, RAG, attribution, and query-rewriting research into a measurement framework for practical AI visibility systems.

\section{A Multistage View of Generative Visibility}
Let $R$ denote a request state, $e$ an engine, $t$ time, and $p$ a page. For a standalone prompt benchmark, $R$ can be approximated by the prompt text. In multi-turn use, prior turns may be part of $R$ \citep{tannenbaum2026prompt,tannenbaum2026context}.

A minimal decomposition of observed citation visibility is
\begin{equation}
P(C=1\mid p,R,e,t)=P(E=1\mid p,R,e,t)\,P(C=1\mid E=1,p,R,e,t),
\label{eq:twostage}
\end{equation}
where $E$ is exposure to the generation or citation process and $C$ is observed citation. The decomposition is intentionally architecture-agnostic. Exposure may depend on search triggering, query rewriting, fan-out, crawling and indexing, retrieval, reranking, source filtering, context allocation, or tool execution. Conditional citation may depend on topical relevance, source reliability, evidence density, context position, answer style, generation behavior, and citation formatting.

A deterministic engine-free score $g(p,R)$ can be useful without estimating either probability exactly. It may estimate content quality, request fit, or a robust proxy for conditional selection. Interpreting $g$ as end-to-end visibility, however, requires an additional bridge: exposure must either be approximately invariant for the comparison being made or be predictable from variables encoded in $g$. Fixed-candidate designs set $E=1$ by construction and therefore identify the second factor rather than the first.

\paragraph{Non-identifiability without an exposure model.} Suppose a controlled experiment identifies a conditional selection function $s(p,R)=P(C=1\mid E=1,p,R)$ perfectly. End-to-end visibility is still not identified from $s$ alone. For any admissible exposure function $e(p,R)\in[0,1]$, the observable citation probability is $e(p,R)s(p,R)$. Two production engines can therefore share the same conditional preference over exposed pages and yet produce arbitrarily different end-to-end citation distributions if their exposure functions differ. Conversely, similar final citation rates can conceal different exposure and conditional-selection mechanisms. This simple product structure is why a fixed-candidate citation experiment and a live-engine visibility audit are complementary rather than interchangeable.

A fuller visibility vector is useful in practice:
\begin{equation}
V(p,R,e,t) = (Q, E, S, A, B, Y),
\end{equation}
where $Q$ is query-page fit, $E$ is exposure, $S$ is citation selection conditional on exposure, $A$ is citation absorption or substantive use in the answer, $B$ is observed brand/recommendation prominence, and $Y$ is a downstream user outcome. Different experiments identify different coordinates of this vector. Treating them as interchangeable creates apparent contradictions that are often only estimand mismatches.

\section{Related Work}
\subsection{Generative Engine Optimization and engine-free scoring}
The foundational GEO work of \citet{aggarwal2024} formalized visibility optimization in generative answers and reported improvements from several content interventions in a controlled evaluation. Since then, GEO has expanded into a heterogeneous literature covering source selection, citation, content structure, competitive effects, and measurement design. \citet{martinez2026} surveys 45 studies from 2023--2026 and explicitly argues that GEO is a stochastic, partially observable pipeline rather than a single ranking task. \citet{allaham2026} audits ChatGPT, Copilot, Gemini, and Perplexity on 712 real-world queries and finds both repeated source concentration and a large tail of minimally cited domains. \citet{yang2025} analyzes more than 366,000 citations from OpenAI, Perplexity, and Google systems in the AI Search Arena and likewise finds provider-specific source choices alongside shared concentration patterns. These audits are particularly relevant to the present paper because they treat the engine's surfaced sources as an outcome to be measured, rather than assuming a common candidate pool. \citet{chen2025geo} compares traditional and AI search across verticals, languages, and paraphrases, reporting differences in source mix, freshness, and sensitivity to phrasing.

\citet{bajemon2026} is most directly relevant here. Their deterministic score is designed to be engine-free and manipulation-resistant. The paper finds weak predictive power for generic page scoring but a much stronger relationship once the request is incorporated. This is consistent with competitive controlled studies such as \citet{vishwakarma2026}, which inject exactly two candidate documents and find topical relevance and context position among the strongest predictors of which source is cited first. Both designs are valuable precisely because they control the candidate set. They therefore do not identify organic exposure in live systems.

\subsection{Production generative search and source divergence}
Multiple audits suggest that live generative-search products differ materially in retrieval and citation behavior. \citet{liu2023} evaluates four generative search engines and finds substantial citation-support gaps, motivating separate measures for citation completeness and citation correctness. \citet{li2024arbiters} audits ChatGPT, Bing Chat, and Perplexity over seven days and reports commercial and geographic biases in source authority. \citet{kirsten2026} compares traditional search with four generative-search engines and documents differences in source breadth, reliance on external retrieval, and surfaced concepts.

Recent large-scale work makes the retrieval-layer distinction especially explicit. Using 11,500 queries, \citet{grossman2026} compares Google organic results, AI Overviews, and Gemini, reporting average source-set Jaccard similarity below 0.2 and reduced consistency across repeated runs. \citet{strauss2025} studies roughly 14,000 search-enabled LMArena conversations and distinguishes pages apparently visited from pages actually cited, documenting a substantial attribution gap that varies by model. These results support the premise that observed citation is a downstream output of engine-specific search and attribution choices rather than a direct reading of page quality.

\subsection{Query rewriting, multi-query retrieval, and candidate-set construction}
The exposure term in Equation~\ref{eq:twostage} is not a single retrieval call. Query rewriting is a standard way to adapt a user request to a retriever. \citet{ma2023rewrite} proposes a rewrite-retrieve-read pipeline in which an LLM generates search queries before web retrieval. \citet{zhang2024adaptive} trains conversational query rewriters against retriever preferences, illustrating that the representation sent to retrieval can materially differ from the user's surface form. Multi-query systems go further: RAG-Fusion \citep{rackauckas2024} and diverse multi-query rewriting \citep{li2024dmqr} intentionally generate multiple retrieval views of the same request to increase document coverage.

These methods also show why retrieval gains need not propagate mechanically to final answers. \citet{lee2025setr} argues that RAG retrieval should be treated as set selection rather than only ranking. In a production-style evaluation, \citet{medrano2026} finds that multi-query fusion increases raw recall but that reranking and context-budget constraints can erase those gains. Thus, even when the search layer expands the candidate pool, later stages can compress it again.

\subsection{Citation selection, attribution, and absorption}
Citation is not equivalent to source use. \citet{gao2023} studies language-model generation with citations, while \citet{nematov2025} adapts attribution methods to estimate the influence of retrieved documents in RAG. \citet{zhang2026} explicitly separates citation selection from citation absorption, showing that engines can differ in citation breadth and the degree to which cited pages contribute to generated answers. \citet{mody2026} similarly shows that compression can preserve surface citation precision while weakening re-attribution to original source spans. These works motivate treating exposure, citation, and substantive answer influence as separate outcomes.

\subsection{Conversation state and downstream behavior}
Our four prior papers extend the same distinction beyond retrieval. \citet{tannenbaum2026arsd} formalizes answer-reconstruction search density, quantifying the search work compressed into answers. \citet{tannenbaum2026prompt} shows that the final user message is often not a complete representation of the active request state. \citet{tannenbaum2026context} experimentally restores missing within-conversation context and measures its effect on answers. \citet{tannenbaum2026purchase} separates visible recommendation behavior from later buyer response. Together with the present study, these results suggest that robust AI-search measurement requires explicit boundaries around request state, retrieval state, answer state, and user outcome.

\section{Data and Design}
\subsection{Benchmark and collection window}
We use an internal Aiso citation-monitoring benchmark consisting of 15 English-language commercial prompts about AI search visibility software. The benchmark was run in a United States locale against production AI search surfaces. The stored data contain the prompt, service, cited page title, exact URL string, citation position, collection date, and extracted domain. No user-level personal data are used in this study.

The full collection contains 1,435 citation rows from 4--6 June 2026. The 4 June and 5 June snapshots are exact duplicates at the prompt-engine citation-set level. Because this could reflect caching, collection reuse, or genuinely identical outputs, we do not treat that transition as independent temporal evidence. Cross-engine analysis therefore uses the 6 June snapshot only. Temporal analysis uses only the 5 June to 6 June transition.

On 6 June, the dataset contains 589 citation observations: 231 from ChatGPT, 63 from Copilot, 145 from Google, and 150 from Perplexity. These correspond to 528 distinct stored URL strings and 356 distinct extracted domains. Ten prompts are present on all four engines; the remaining prompts are present on at least two.

\begin{table}[t]
\centering
\caption{6 June 2026 benchmark coverage. Distinct counts use exact stored URL strings and extracted domains.}
\label{tab:coverage}
\begin{tabular}{lrrrr}
\toprule
Engine & Citation rows & Prompts & Unique URLs & Unique domains \\
\midrule
ChatGPT & 231 & 13 & 201 & 155 \\
Copilot & 63 & 14 & 62 & 52 \\
Google & 145 & 12 & 136 & 95 \\
Perplexity & 150 & 15 & 148 & 105 \\
\midrule
All engines & 589 & 15 & 528 & 356 \\
\bottomrule
\end{tabular}
\end{table}

\subsection{Primary overlap outcomes}
For each prompt $q$, engine $e$, and date $t$, let $S_{qet}$ denote the set of cited exact URLs and $D_{qet}$ the set of cited domains. For two engines $e_1,e_2$ on the same prompt and date, exact-URL overlap is
\begin{equation}
J_U(q,e_1,e_2,t)=\frac{|S_{qe_1t}\cap S_{qe_2t}|}{|S_{qe_1t}\cup S_{qe_2t}|}.
\end{equation}
Domain overlap $J_D$ is defined analogously. The primary analysis reports pairwise means, medians, zero-overlap shares, and prompt-clustered bootstrap confidence intervals.

Jaccard can be small simply because one engine returns more citations than another. We therefore also report containment,
\begin{equation}
K_U(A,B)=\frac{|A\cap B|}{\min(|A|,|B|)},
\end{equation}
which asks whether the smaller set is largely contained in the larger set. On the four-engine-complete subset, mean containment is 0.021, so unequal set sizes do not explain the result.

\subsection{Rank-sensitive outcomes}
Citation position may carry different interface semantics across products, so exact sets remain our primary outcome. As robustness checks, we compute top-5 and top-10 exact-URL Jaccard, plus reciprocal-rank-weighted Jaccard. For URL $u$ in engine output $A$, define weight $w_A(u)=1/r_A(u)$ using its recorded citation position. Weighted Jaccard is
\begin{equation}
J_w(A,B)=\frac{\sum_u \min(w_A(u),w_B(u))}{\sum_u \max(w_A(u),w_B(u))}.
\end{equation}
If different engines agree on a prominent core but diverge only in the tail, these rank-sensitive measures should be substantially higher than unweighted full-set comparisons.

\subsection{Single-engine proxy coverage and union expansion}
For the ten prompts observed on all four engines, let $U_q=\bigcup_e S_{qe}$ be the observed four-engine URL union. For each engine we report $|S_{qe}|/|U_q|$, the fraction of that multi-engine union captured by monitoring only one engine. We also define union expansion as
\begin{equation}
X_q=\frac{|U_q|}{\max_e |S_{qe}|},
\end{equation}
which measures how much larger the multi-engine source universe is than the broadest single-engine output.

\subsection{Source exclusivity, breadth, and concentration}
We classify each distinct URL and domain by the number of engines on which it appears on 6 June. We also report mean URLs and domains per prompt by engine. At the aggregate engine level, domain concentration is summarized using a Herfindahl-Hirschman index over citation shares; its reciprocal is reported as an intuitive ``effective number of domains.'' These quantities are descriptive and are not normalized for product market share or traffic.

\subsection{Temporal turnover}
For temporal instability within the same engine, URL turnover from $t_1$ to $t_2$ is
\begin{equation}
T_U(q,e;t_1,t_2)=1-J(S_{qet_1},S_{qet_2}).
\end{equation}
We report full-set, top-5, top-10, and reciprocal-rank-weighted turnover for the 5-to-6 June transition. Google is absent from the earlier snapshot and is therefore excluded from temporal comparisons.

\subsection{Matched-size random-overlap baseline}
Raw Jaccard overlap can be mechanically affected by list length. We therefore add a prompt-specific matched-size baseline for the ten prompts observed on all four engines. For a prompt with observed four-engine URL union size $U$, and two engines returning $m$ and $n$ unique URLs, we consider the overlap that would arise if each engine drew a subset of the same size uniformly without replacement from that observed union. The intersection $X$ then follows a hypergeometric distribution,
\begin{equation}
P(X=x)=\frac{\binom{m}{x}\binom{U-m}{n-x}}{\binom{U}{n}},
\end{equation}
and we compute the exact expected Jaccard $E[X/(m+n-X)]$ and $P(X=0)$. This is a descriptive randomization baseline, not a model of how engines actually select sources. Its purpose is narrower: to test whether near-zero overlap is merely a mathematical consequence of unequal source-list sizes inside the source universe that the four engines collectively exposed.

\subsection{Inference and robustness}
For the principal cross-engine quantities, the resampling unit is the prompt. In each of 10,000 bootstrap replicates, prompts are sampled with replacement and all engine-pair comparisons belonging to a sampled prompt are retained together. This preserves within-prompt dependence. Percentile 2.5th and 97.5th quantiles form the reported intervals. The same prompt-cluster procedure is used for temporal turnover. Because the benchmark contains only 15 prompts, these intervals characterize finite-sample uncertainty inside this prompt set; they do not support population-level claims across all search tasks.

\section{Results}
\subsection{Cross-engine exact-URL overlap is nearly zero}
Across 73 same-prompt engine pairs on 6 June, mean exact-URL Jaccard overlap is 0.0079 (95\% cluster-bootstrap CI 0.0037--0.0132) and the median is 0. Fully 84.9\% of pairwise comparisons share no exact URL at all (95\% CI 75.4\%--93.2\%). Mean domain Jaccard overlap is higher but still small at 0.0265 (95\% CI 0.0176--0.0375), with 60.3\% of pairs sharing no domain.

\begin{figure}[t]
\centering
\includegraphics[width=0.88\linewidth]{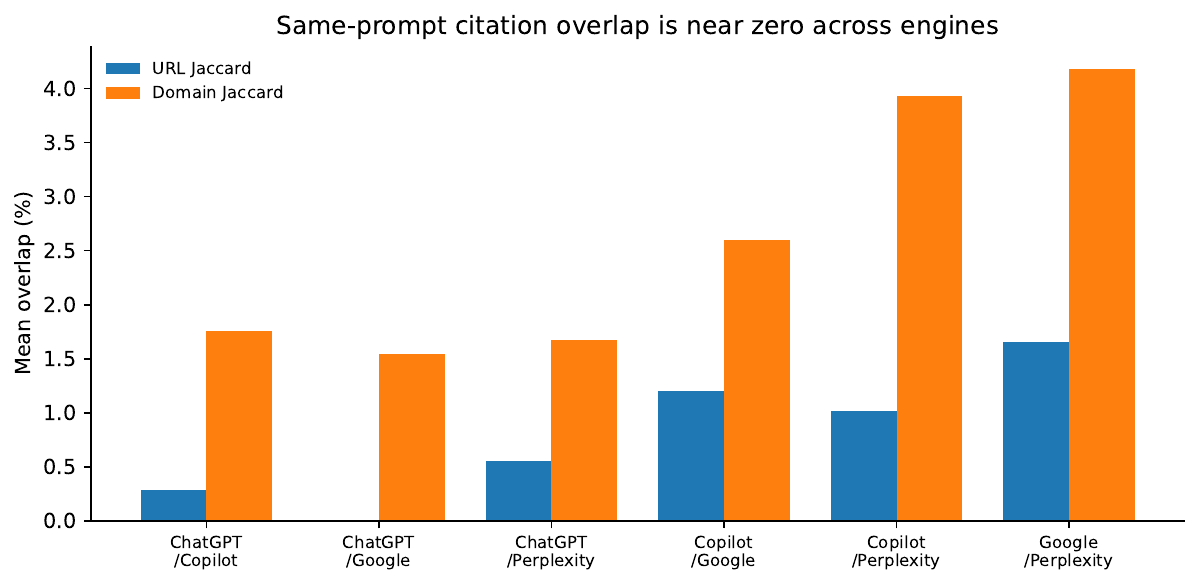}
\caption{Mean pairwise Jaccard overlap of cited exact URLs and domains for the same prompt on 6 June 2026.}
\label{fig:overlap}
\end{figure}

Figure~\ref{fig:overlap} shows that this pattern is not driven by a single engine pairing. The largest mean exact-URL overlap, Google versus Perplexity, is only 0.0165. ChatGPT and Google share no exact cited URL in any of their 11 overlapping prompt comparisons.

\begin{table}[t]
\centering
\caption{Pairwise same-prompt overlap on 6 June 2026.}
\label{tab:pairs}
\begin{tabular}{lrrrr}
\toprule
Engine pair & $n$ prompts & URL Jaccard & Domain Jaccard & Zero URL overlap \\
\midrule
ChatGPT / Copilot & 12 & 0.0029 & 0.0176 & 91.7\% \\
ChatGPT / Google & 11 & 0.0000 & 0.0154 & 100.0\% \\
ChatGPT / Perplexity & 13 & 0.0055 & 0.0167 & 84.6\% \\
Copilot / Google & 11 & 0.0120 & 0.0260 & 81.8\% \\
Copilot / Perplexity & 14 & 0.0102 & 0.0393 & 85.7\% \\
Google / Perplexity & 12 & 0.0165 & 0.0418 & 66.7\% \\
\bottomrule
\end{tabular}
\end{table}

\subsection{Agreement does not reappear at the top of the citation list}
A simple explanation for low full-set overlap would be that engines share a common core of prominent sources and disagree only in the tail. The rank-sensitive analysis does not support that explanation in this benchmark.

Restricting to the ten prompts observed on all four engines yields exactly 60 prompt-engine-pair comparisons. Mean full-set exact-URL Jaccard is 0.0072. Mean reciprocal-rank-weighted Jaccard falls to 0.0027. Mean top-10 Jaccard is 0.0066. Most strikingly, top-5 exact-URL Jaccard is exactly zero across all 60 comparisons: no pair of engines shares an exact URL within their respective first five recorded citation positions on any of the ten complete prompts.

The result is robust to unequal list lengths. Mean containment $|A\cap B|/\min(|A|,|B|)$ is only 0.021. Thus, low Jaccard is not simply an artifact of one engine citing many more pages than another.

\subsection{A single engine is a weak proxy for the multi-engine source universe}
On the ten four-engine-complete prompts, the union contains a mean of 43.4 exact URLs per prompt. The broadest single-engine set contains 18.8 on average. The observed four-engine union is therefore 2.43 times as large as the broadest single engine.

No engine captures half of the observed four-engine URL union. ChatGPT has the highest mean union recall at 42.6\%, followed by Google at 24.3\%, Perplexity at 23.6\%, and Copilot at 11.4\%.

\begin{figure}[t]
\centering
\includegraphics[width=0.72\linewidth]{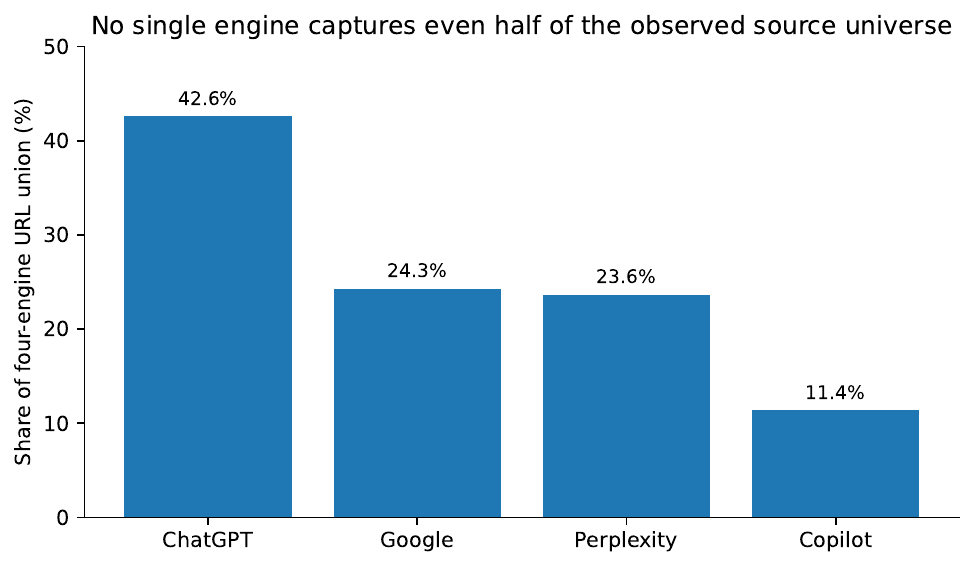}
\caption{Mean fraction of the four-engine exact-URL union captured by each single engine, restricted to the ten prompts observed on all four engines. These are source-coverage statistics, not estimates of user traffic or market share.}
\label{fig:proxy}
\end{figure}

Operationally, this means that a measurement system observing one engine and silently treating it as representative of ``AI search'' would miss most of the sources present elsewhere in this benchmark even before considering answer wording, brand mentions, recommendation order, or user behavior.

\subsection{Low overlap is far below a matched-size random baseline}
On the ten four-engine-complete prompts, the matched-size hypergeometric baseline implies mean expected exact-URL Jaccard of 0.1272. The observed value is 0.0072, only 5.7\% of that baseline. Under the same baseline, the mean probability of zero overlap is 12.3\%; the observed zero-overlap share is 86.7\%.

The gap appears in every engine pair. For ChatGPT versus Google, for example, the matched-size baseline implies mean expected Jaccard of 0.1793 and mean zero-overlap probability below 1\%, yet the observed Jaccard is 0 and all ten matched prompts have zero shared exact URL. ChatGPT versus Perplexity has expected Jaccard 0.1793 versus observed 0.0029. Even the most overlapping pair, Google versus Perplexity, has observed Jaccard 0.0165 against a matched-size expectation of 0.1359.

\begin{figure}[t]
\centering
\includegraphics[width=0.88\linewidth]{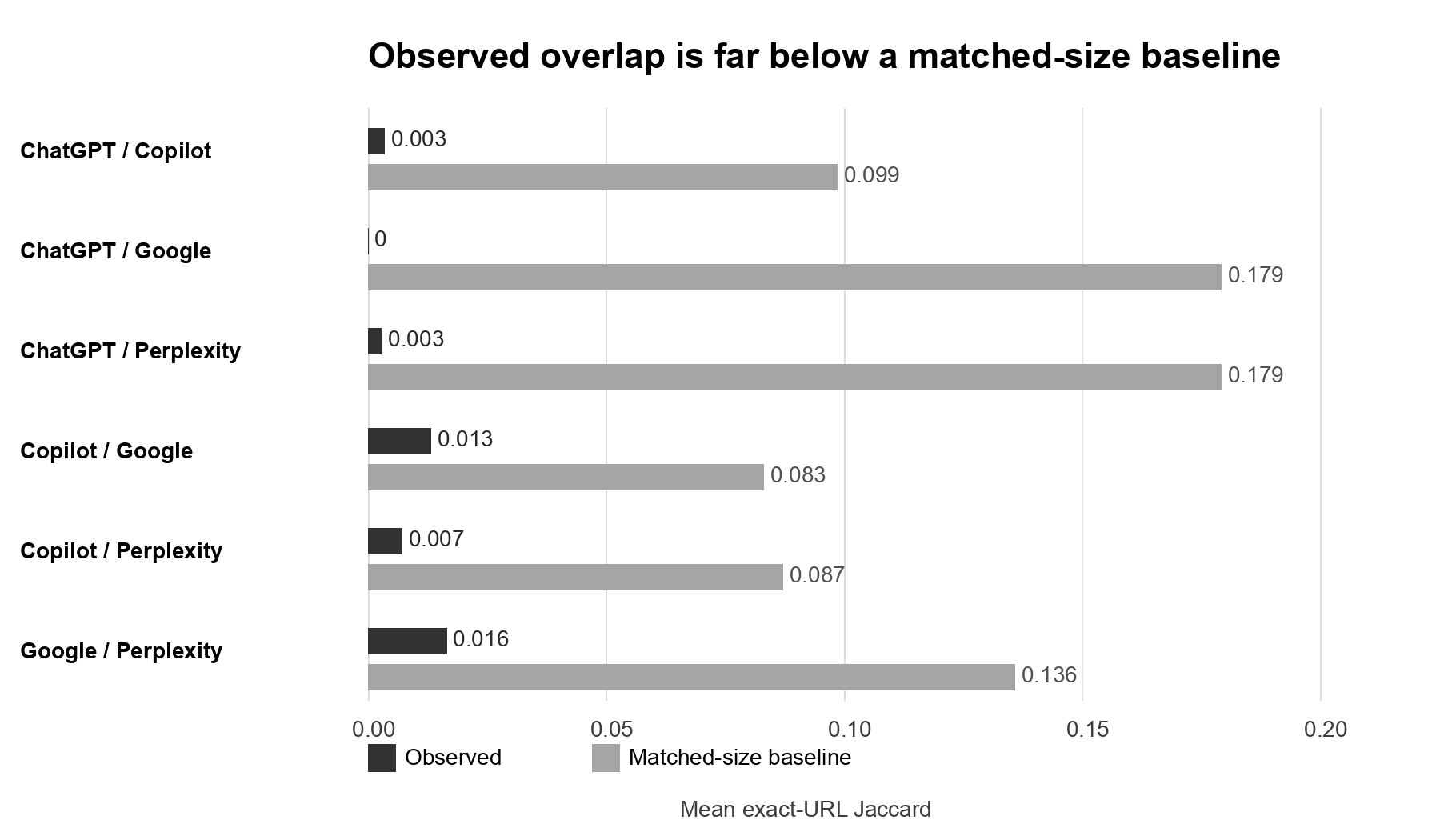}
\caption{Observed exact-URL Jaccard versus the prompt-specific matched-size random-overlap baseline on ten prompts observed on all four engines. The baseline preserves each prompt's four-engine URL universe and each engine's number of cited URLs.}
\label{fig:random}
\end{figure}

This comparison does not imply that production engines should behave like uniform random samplers. It shows something more limited but useful: list-length differences cannot explain the observed near-disjoint citation sets. The engines partition the jointly observed source universe much more strongly than a size-matched random allocation would. In absolute terms, each engine also adds substantial material that none of the other three engines surfaced on the same prompt: a mean of 18.6 exact URLs for ChatGPT, 10.2 for Google, 9.5 for Perplexity, and 4.3 for Copilot in the complete subset.

\subsection{Most observed sources are engine-specific}
Among the 528 distinct exact URLs observed on 6 June, 509 appear in only one engine. That is 96.4\% of all observed URLs. Nineteen URLs appear in two engines; none appears on all four. At domain level, 317 of 356 domains (89.0\%) appear in one engine only. Twenty-eight appear in two engines, ten in three engines, and one domain appears in all four.

The per-engine picture is similar in the four-engine-complete subset. On average, 99.2\% of ChatGPT's cited URLs, 92.1\% of Copilot's, 94.5\% of Google's, and 95.0\% of Perplexity's are unique to that engine within the matched prompt source universe.

\begin{figure}[t]
\centering
\includegraphics[width=0.72\linewidth]{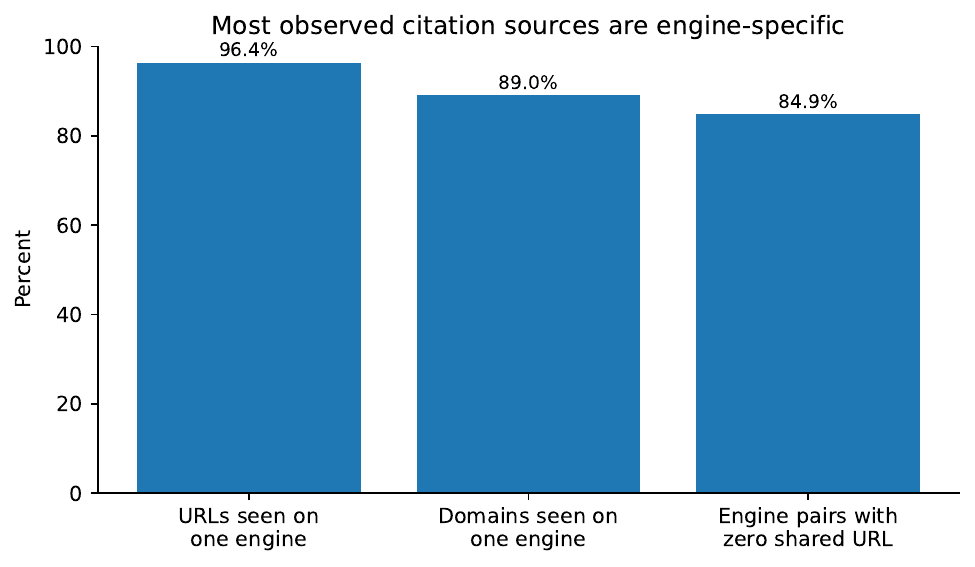}
\caption{Engine specificity in the 6 June snapshot. The first two bars classify unique observed URLs and domains by whether they appear on only one engine. The third bar is the share of same-prompt engine-pair comparisons with zero shared exact URL.}
\label{fig:specificity}
\end{figure}

\subsection{Engines also differ in citation breadth and source concentration}
Low overlap could coexist with identical citation breadth, but the engines also return materially different numbers of sources per prompt. ChatGPT cites a mean of 17.8 exact URLs per observed prompt on 6 June, Google 12.1, Perplexity exactly 10.0 in this extract, and Copilot 4.5. Domain breadth follows the same ordering, though less sharply.

\begin{figure}[t]
\centering
\includegraphics[width=0.68\linewidth]{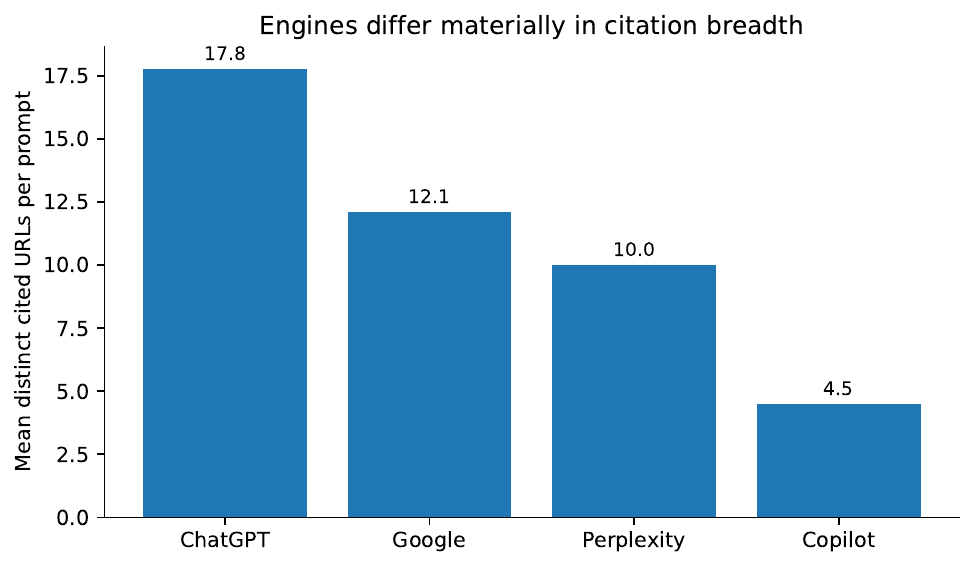}
\caption{Mean number of distinct exact cited URLs per prompt on 6 June 2026.}
\label{fig:breadth}
\end{figure}

\begin{table}[t]
\centering
\caption{Citation breadth and aggregate domain concentration on 6 June. ``Effective domains'' is the reciprocal of the domain-share HHI.}
\label{tab:breadth}
\begin{tabular}{lrrrr}
\toprule
Engine & URLs/prompt & Domains/prompt & Effective domains & Top-10 domain share \\
\midrule
ChatGPT & 17.8 & 15.2 & 60.6 & 28.6\% \\
Copilot & 4.5 & 4.2 & 43.6 & 33.3\% \\
Google & 12.1 & 10.3 & 42.5 & 33.8\% \\
Perplexity & 10.0 & 8.4 & 48.3 & 30.7\% \\
\bottomrule
\end{tabular}
\end{table}

These differences matter for benchmarking because raw citation counts, share-of-citations, and source-diversity metrics are partly properties of the product's citation policy. A page that is never eligible for a four-source answer has a different opportunity set from a page evaluated in an 18-source answer.

\subsection{Cross-engine divergence is broad across prompts}
The headline result is not produced by one or two anomalous prompts. Five prompts have mean pairwise exact-URL overlap of zero across the engines on which they are observed. The most convergent prompt reaches only 0.0284 mean pairwise Jaccard. The four-engine-complete subset yields a mean URL Jaccard of 0.0072 and zero exact-URL overlap in 86.7\% of pairwise comparisons, almost identical to the full matched-pair analysis. Uneven engine coverage is therefore not the primary explanation.

\subsection{Same-engine sets move over time, but engine identity dominates this comparison}
The exact duplicate 4 June and 5 June snapshots are excluded from temporal inference. Between 5 and 6 June, 41 same-prompt pairs are available across ChatGPT, Copilot, and Perplexity. Mean exact-URL-set turnover is 67.0\% (95\% prompt-clustered bootstrap CI 61.1\%--72.2\%). Reciprocal-rank-weighted turnover is 69.3\%; top-5 turnover is 68.9\%; top-10 turnover is 66.8\%.

By engine, full-set turnover is 82.6\% for ChatGPT, 76.6\% for Copilot, and 45.5\% for Perplexity. The prompt count is small and the transition may reflect collection-system changes as well as engine drift, so these values should not be generalized as daily volatility estimates.

\begin{figure}[t]
\centering
\includegraphics[width=0.70\linewidth]{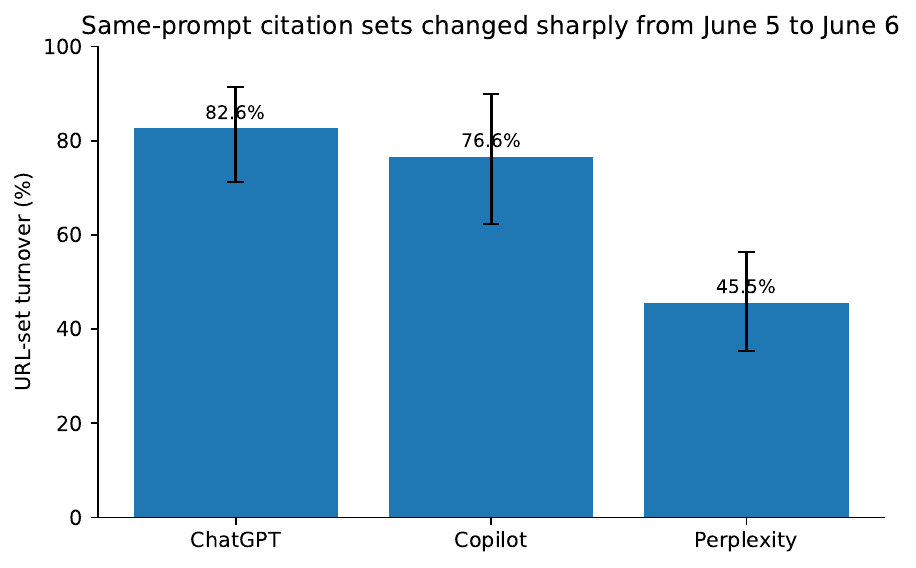}
\caption{Same-prompt exact-URL-set turnover from 5 to 6 June 2026. Error bars are prompt-clustered bootstrap 95\% intervals. Google is omitted because it has no 5 June observations in the extract.}
\label{fig:turnover}
\end{figure}

The comparison is nevertheless informative. Mean same-engine adjacent-day URL Jaccard is $1-0.670=0.330$, versus 0.0079 for same-day cross-engine comparisons. In this benchmark, an adjacent-day source set from the \emph{same} engine is therefore about 42 times more similar than a same-day source set from a \emph{different} engine. At domain level, the analogous same-engine adjacent-day similarity is about 0.368 versus 0.0265 across engines, roughly a 14-fold difference. Even with substantial short-horizon movement, engine identity is the larger source of observable citation-set separation here.

\section{Reinterpreting ``Scoring Without the Engine''}
The central empirical result in \citet{bajemon2026} is not contradicted by our analysis. The two studies fit naturally inside Equation~\ref{eq:twostage}. Their controlled candidate-set experiments hold exposure fixed by supplying the sources. The design can therefore evaluate how much page quality or query-page fit predicts conditional citation. Our benchmark does not observe that latent candidate set. It measures the sources that survive to visible production citations after each engine has executed its own search, retrieval, ranking, generation, and attribution process.

The studies therefore address distinct estimands:
\begin{enumerate}[leftmargin=*]
\item \textbf{Engine-free page quality}: a deterministic property of a page.
\item \textbf{Query-page fit}: relevance or usefulness conditional on a request representation.
\item \textbf{Engine exposure}: whether a page becomes available to a production answer process.
\item \textbf{Conditional citation selection}: whether the page is cited given exposure.
\item \textbf{Citation absorption}: whether the source materially contributes evidence or language to the answer \citep{zhang2026,nematov2025}.
\item \textbf{Observed answer visibility}: final citation, brand mention, recommendation, or prominence.
\item \textbf{User outcome}: subsequent action, which may not be observable from the answer or conversation record \citep{tannenbaum2026purchase}.
\end{enumerate}

An engine-free score can legitimately target Items 1 or 2. Fixed-context experiments can target Item 4 and sometimes Item 5. Live engine monitoring is needed to observe Items 3 and 6. Item 7 requires a separate behavioral measurement design.

The low cross-engine overlap makes one interpretation difficult to sustain without additional evidence: that a single engine-free score can be read directly as a stable probability of citation across production engines. Such a claim requires either a model of engine exposure or evidence that exposure is sufficiently invariant for the intended use case. In this benchmark, observable exposure outcomes are strongly engine-specific.

\section{Implications for GEO Measurement Systems}
\subsection{Report a vector, not one undifferentiated score}
A practical evaluation system should distinguish at least query-page fit, observed engine exposure, and observed citation or answer influence. A single scalar may still be useful for interface simplicity, but the underlying components should remain recoverable. Otherwise a change in retrieval exposure can be mistaken for a change in page quality, and vice versa.

\subsection{Do not use one engine as an unlabelled universal control}
The proxy-coverage analysis shows why single-engine monitoring can be misleading. In the four-engine-complete subset, even the broadest engine captures less than half of the union on average. This does not imply that every use case requires four engines. It does imply that engine choice is a sampling decision and should be visible in the metric definition.

\subsection{Prominence does not rescue agreement in this benchmark}
Top-five exact-URL overlap is zero in the complete subset, and reciprocal-rank-weighted overlap is lower than unweighted overlap. Researchers should not assume that prominent citations form a shared core across engines. Because citation-position semantics differ by product surface, rank-sensitive metrics should be presented as supplementary rather than silently pooled.

\subsection{Repeated measurement is part of the estimand}
A live citation result is a draw from a changing system. Longitudinal designs should report run frequency, dates, locale, product mode, search setting, caching controls, and how repeated runs are aggregated. The exact duplicate 4-to-5 June snapshot in our own data demonstrates why this metadata matters: repeated rows do not automatically imply independent repeated observations.

\subsection{Query representation remains a separate validity threat}
This study uses standalone benchmark prompts, but real conversational search can distribute the request across prior turns. Query rewriting research shows that retrieval effectiveness can change when a conversational request is decontextualized or expanded \citep{qian2022,ma2023rewrite,ye2023,mo2023,zhang2024adaptive,choi2024}. Our prior work similarly finds that the endpoint prompt often omits active constraints \citep{tannenbaum2026prompt}, and controlled regeneration shows that restoring those constraints changes answers \citep{tannenbaum2026context}. Engine-specific monitoring built on isolated endpoint prompts can therefore still misrepresent the request users actually made.

\subsection{Citation should not be treated as synonymous with influence or trust}
Even a displayed citation may not indicate that a source materially supported the answer. Citation-support studies find substantial gaps between fluent answers and verifiable sourcing \citep{liu2023}. Citation-absorption and attribution studies reinforce the need to inspect whether source content actually contributed to claims \citep{qi2024,nematov2025,abolghasemi2025,vericite2025,zhang2026,mody2026}. At the user layer, \citet{li2025trust} finds that citations can increase trust even when incorrect, making citation presence both a visibility outcome and a user-interface intervention. These are different objects and should be measured separately.

\section{Threats to Validity and Limitations}
\paragraph{Narrow domain.} The 15 prompts concern AI visibility software. This is a commercially relevant but unusually self-referential category in which vendors, marketing publications, SEO platforms, and comparison pages may be overrepresented. The numerical overlap rates should not be generalized to travel, health, shopping, local search, news, or other verticals without replication.

\paragraph{Small prompt set.} The primary snapshot contains only 15 prompts, ten of which are available on all four engines. Prompt-clustered bootstrap intervals reflect sampling variation within this benchmark, not uncertainty over all possible information needs.

\paragraph{Displayed citations are not retrieval traces.} We observe final displayed citations. We do not observe hidden search queries, candidate pages, reranker scores, context windows, source reads, or model internals. Low citation overlap therefore cannot identify the causal stage responsible for divergence. This is why we use the term \emph{observable citation set} rather than \emph{retrieval set}. \citet{strauss2025} provides complementary evidence that visited and cited sources can differ substantially.

\paragraph{Exact URL identity is strict.} Different URLs may host duplicate, syndicated, canonicalized, translated, or semantically equivalent information. Domain-level overlap partially relaxes this issue and remains low, but domain equality is also an imperfect semantic measure. Future work should add canonical URL resolution and content-level equivalence clustering.

\paragraph{Product-surface comparability.} Citation positions and interface behavior may not mean exactly the same thing across ChatGPT, Copilot, Google, and Perplexity. We use exact sets as the primary outcome because they make fewer assumptions about cross-product ranking semantics. Rank-sensitive analyses are robustness checks.

\paragraph{Temporal ambiguity.} The 4 June and 5 June sets are exact duplicates. The 5-to-6 June shift could reflect engine change, collection change, or both. We therefore describe temporal turnover as a diagnostic, not an estimate of daily volatility.

\paragraph{No causal content intervention.} This paper does not edit pages and then measure live retrieval changes. It therefore cannot show that any specific content modification causes an exposure increase. The evidence is about the measurement boundary between page scoring and observed production citation, not about a particular optimization tactic.

\paragraph{Commercial conflict of interest.} The author is the founder of Aiso Boost Ltd., which develops software for measuring and improving visibility in AI search. The benchmark was collected through Aiso infrastructure. This creates a direct commercial conflict of interest. The statistical endpoints are transparent set-based quantities, and the source package releases aggregate values and figure-generation code, but the full operational citation log is not released. Independent replication on public prompt sets and independently collected engine outputs is necessary.

\section{Future Work}
The most important next step is a larger repeated-run panel spanning verticals, languages, and request types. Each prompt-engine cell should be executed independently several times per day, with product mode, search activation, locale, and model version recorded where available. This would permit variance decomposition into prompt, engine, run, and time components rather than relying on two snapshots.

A second extension should pair observable citations with search or retrieval traces whenever platforms expose them. This would allow Equation~\ref{eq:twostage} to be estimated directly: first model exposure, then model conditional citation. A third extension should score candidate pages offline and test whether high query-page-fit pages are actually surfaced by each engine. That design would directly measure the bridge from engine-free scoring to production visibility rather than evaluating the stages separately.

A fourth extension should canonicalize pages semantically. Exact URL matching is intentionally conservative; a content-fingerprint layer could identify mirrored, syndicated, translated, or near-duplicate pages and estimate overlap at document-cluster level. A fifth should restore conversational request state and compare endpoint-prompt scoring with conversation-conditioned scoring, connecting the retrieval question in this paper with our prior context work \citep{tannenbaum2026prompt,tannenbaum2026context}.

Finally, end-to-end GEO experiments should manipulate pages and measure not only final citation but intermediate outcomes: search triggering, fan-out queries, retrieved candidate presence, citation, absorption, brand mention, recommendation position, click behavior, and conversion where ethically and practically observable. Without this decomposition, a successful intervention cannot be localized to the stage it actually changed.

\section{Conclusion}
Engine-free scoring can be useful, reproducible, and cheap. The mistake is to give it a broader interpretation than the design supports. In this benchmark, production engines do not converge on a stable shared citation set. Mean same-prompt exact-URL overlap is below 1\%, most engine pairs share no exact URL, top-five overlap is zero on the complete four-engine subset, more than 96\% of observed URLs appear on only one engine, and no single engine captures even half of the four-engine source union. Source sets also change substantially across the one non-duplicate temporal transition we can evaluate.

The synthesis with \citet{bajemon2026} is therefore straightforward. A deterministic score without the engine can characterize page quality or query-page fit. A controlled candidate-set experiment can estimate conditional citation preference. Production visibility additionally depends on engine-specific exposure, and final answer influence is another stage again. Treating these as separate quantities makes GEO evaluation more precise and makes apparently conflicting results easier to reconcile.

\section*{Data and Code Availability}
The arXiv source package includes the aggregate table used for the figures and a Python script that regenerates the figures from those aggregates. The underlying operational citation log is not released in full because it is part of an internal monitoring dataset. The released aggregates are sufficient to reproduce the figures and descriptive quantities in this version but are not sufficient to independently audit raw extraction or URL canonicalization.

\section*{Acknowledgments}
The author thanks the Aiso team for maintaining the citation-monitoring infrastructure used in this audit.

\appendix
\section{Four-engine-complete subset}
\begin{table}[h]
\centering
\caption{Per-engine proxy coverage and source uniqueness. Values are means across ten prompts observed on all four engines.}
\begin{tabular}{lrrr}
\toprule
Engine & URLs per prompt & Recall of four-engine union & Own URLs unique to engine \\
\midrule
ChatGPT & 18.8 & 42.6\% & 99.2\% \\
Copilot & 4.7 & 11.4\% & 92.1\% \\
Google & 10.7 & 24.3\% & 94.5\% \\
Perplexity & 10.0 & 23.6\% & 95.0\% \\
\bottomrule
\end{tabular}
\end{table}

\begin{table}[h]
\centering
\caption{Rank-sensitive overlap on ten prompts observed on all four engines.}
\begin{tabular}{lr}
\toprule
Metric & Mean across 60 engine-prompt pairs \\
\midrule
Exact URL Jaccard & 0.0072 \\
Reciprocal-rank-weighted URL Jaccard & 0.0027 \\
Top-5 exact URL Jaccard & 0.0000 \\
Top-10 exact URL Jaccard & 0.0066 \\
Containment $|A\cap B|/\min(|A|,|B|)$ & 0.0210 \\
Zero exact URL overlap & 86.7\% \\
\bottomrule
\end{tabular}
\end{table}

\section{Temporal robustness metrics}
Across the 41 prompt-engine pairs observed on both 5 and 6 June, mean full-set URL turnover is 0.670, top-5 turnover is 0.689, top-10 turnover is 0.668, and reciprocal-rank-weighted turnover is 0.693. These similar values indicate that the temporal difference is not isolated to low-ranked citations.

\section{Bootstrap procedure}
For the principal cross-engine quantities, the resampling unit is the prompt. In each of 10,000 bootstrap replicates, prompts are sampled with replacement from the observed prompt set. All available engine-pair comparisons belonging to each selected prompt are included, preserving within-prompt dependence. The statistic is recomputed on the resulting replicate, and percentile 2.5th and 97.5th quantiles form the reported interval. The same prompt-cluster logic is used for the 5-to-6 June temporal turnover intervals.

The bootstrap intervals quantify finite-sample uncertainty within the observed benchmark. They should not be interpreted as confidence intervals for all possible commercial prompts or all generative-search verticals.

\section{Matched-size baseline details}
\begin{table}[h]
\centering
\caption{Observed versus matched-size expected exact-URL Jaccard on the ten prompts observed on all four engines.}
\begin{tabular}{lrrr}
\toprule
Engine pair & Observed Jaccard & Expected Jaccard & Observed zero overlap \\
\midrule
ChatGPT / Copilot & 0.0034 & 0.0986 & 90.0\% \\
ChatGPT / Google & 0.0000 & 0.1793 & 100.0\% \\
ChatGPT / Perplexity & 0.0029 & 0.1793 & 90.0\% \\
Copilot / Google & 0.0132 & 0.0831 & 80.0\% \\
Copilot / Perplexity & 0.0071 & 0.0872 & 90.0\% \\
Google / Perplexity & 0.0165 & 0.1359 & 70.0\% \\
\bottomrule
\end{tabular}
\end{table}

The corresponding mean matched-size probability of zero overlap is 12.3\%, versus an observed zero-overlap share of 86.7\%. This baseline is conditional on the four-engine source union observed for each prompt and preserves the two compared engines' set sizes. It is not intended as a behavioral model of source selection.

\section{Recommended reporting checklist}
For GEO and AI-visibility studies, we recommend reporting the following design choices explicitly:
\begin{itemize}[leftmargin=*]
\item whether candidate sources are fixed by the experimenter or discovered by the engine;
\item whether the request unit is a standalone prompt, a reconstructed request state, or a full conversation;
\item whether the outcome is retrieval/exposure, displayed citation, citation prominence, answer absorption, brand mention, recommendation, or downstream user action;
\item the engine, product surface, locale, date, model or mode when known, and search setting;
\item how URL canonicalization, duplicate pages, syndicated content, and domain extraction are handled;
\item whether repeated runs are independent live executions, cached snapshots, or reused outputs;
\item whether rank-sensitive metrics are comparable across the evaluated interfaces;
\item whether the study has commercial incentives connected to the measured outcome.
\end{itemize}

\end{document}